\documentclass[11pt,english]{iopart} 

\usepackage[usenames,dvipsnames]{color}
\usepackage[normalem]{ulem}
\usepackage{graphicx}

\expandafter\let\csname equation*\endcsname\relax
\expandafter\let\csname endequation*\endcsname\relax

\usepackage{amsmath}
\usepackage{amssymb,amsthm}
\usepackage{amsfonts}
\usepackage{color}
\usepackage{graphics}
\usepackage{epsfig}
\usepackage{bbm}
\usepackage{pstricks}
\usepackage{subfigure}
\usepackage{bm}
\usepackage{bbold}
\usepackage{cite}

\newcommand{\be}{\begin{equation}}
\newcommand{\ee}{\end{equation}}
\newcommand{\bea}{\begin{eqnarray}}
\newcommand{\eea}{\end{eqnarray}}

\newcommand{\mean}[1]{\ensuremath{\langle{#1}\rangle}}

\newcommand{\eins}{\mathbb{1}}

\newcommand{\WW}{\ensuremath{\mathcal{W}}}

\newcommand{\NN}{\ensuremath{\mathcal{N}}}

\newcommand{\HH}{\ensuremath{\mathcal{H}}}
\newcommand{\FF}{\ensuremath{\mathcal{F}}}

\newcommand{\QQ}{\ensuremath{\mathcal{Q}}}

\newcommand{\ketbra}[1]{\ensuremath{| #1 \rangle \!\langle #1 |}}

\newcommand{\ket}[1]{\ensuremath{|#1\rangle}}

\newcommand{\bra}[1]{\ensuremath{\langle#1|}}

\newcommand{\kommentar}[1]{}

\newcommand{\vr}{\ensuremath{\varrho}}

\newcommand{\forget}[1]{}

\begin{document}


\title{Analysing multiparticle quantum states}

\author{Otfried G\"uhne$^1$, Matthias Kleinmann$^2$, and
Tobias Moroder$^1$}
\address{$^1$Naturwissenschaftlich-Technische Fakult\"at,
Universit\"at Siegen,
Walter-Flex-Stra{\ss}e~3,
57068 Siegen, Germany}

\address{$^2$ Department of Theoretical Physics, 
University of the Basque Country UPV/EHU, 
P.O. Box 644, E-48080 Bilbao, Spain
}


\date{\today}


\begin{abstract}
The analysis of multiparticle quantum states is a central problem 
in quantum information processing. This task poses several challenges
for experimenters and theoreticians. We give an overview over current
problems and possible solutions concerning systematic errors of quantum
devices, the reconstruction of quantum states, and the analysis of correlations and complexity in
multiparticle density matrices. 
\end{abstract}




\section{Introduction}
The analysis of quantum states is important for the advances
in quantum optics and quantum information processing. Many experiments
nowadays aim at the generation and observation of certain quantum states 
and quantum effects. For instance, in quantum simulation experiments thermal 
or ground states of certain spin models should be observed. Another 
typical problem is the demonstration of advanced quantum control by 
preparing certain highly entangled states using systems such as 
trapped ions, superconducting qubits, nitrogen-vacancy centers 
in diamond, or polarized photons.

All these experiments require a careful analysis in order to verify
that the desired quantum phenomenon has indeed been observed. This 
analysis does not only concern the final data reported in the experiment 
but in fact, many more questions have to be considered in parallel. Did 
the experimenter align the measurement devices correctly? Have the count 
rates been evaluated properly in order to obtain the mean values of 
the measured observables? Such questions are relevant and, as we demonstrate
below, ideas from theoretical physics can help the 
experimenters answer them. 

\begin{figure}[t]
\begin{center}
\includegraphics[width=0.80\textwidth]{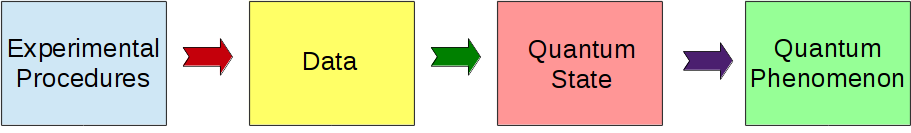}
\end{center}
\caption{The analysis of many experiments in quantum physics can be divided
into several steps, from the experimental procedures to the verification of 
quantum mechanical properties of the generated states.}
\label{exp-scheme}
\end{figure}

Many experiments in quantum optics can be divided in several steps (see also 
Fig.~\ref{exp-scheme}). In the beginning, some experimental procedures are carried out 
and measurements are taken. The results of the measurements are collected as 
data. These data are then processed to obtain a quantum state or density matrix 
$\varrho$, which 
is often viewed as the best description of the ``actual state'' generated in
the experiment. This quantum state can then be further analysed, for instance, 
its entanglement properties may be determined. 

In this article, we show how ideas from statistics and entanglement 
theory can be used for analysing the transitions between the four building blocks
in Fig.~\ref{exp-scheme}. First, we consider the transition from the 
experimental procedures to the data. We show that applying statistical 
tests to the data can be used to recognize systematic errors in the experimental 
procedures, such as a misalignment of the measurement devices. Then, we consider 
the reconstruction of a quantum state from the experimental data. We explain why 
many frequently used state reconstruction schemes, such as the maximum-likelihood 
reconstruction, lead to a bias in the resulting state. This can, for instance, 
result in a fake detection of entanglement, meaning that the reconstructed state 
is entangled, while the original state, on which the measurements were carried out, 
was not entangled. We also show how such a bias can be avoided. Finally, we 
discuss the characterization of quantum states on a purely theoretical level. 
Assuming a multiparticle density matrix $\vr$ we show how its entanglement can 
be  characterized and how the complexity of the state can be quantified using 
tools from information geometry and exponential families.

\section{Systematic errors in quantum experiments}

In this first part of the article, we discuss what assumptions are 
typically used in quantum experiments. The violation of these assumptions
leads to systematic errors and we show how these systematic errors can be
identified using statistical methods and hypothesis tests.

\subsection{Assumptions underlying quantum experiments}
Before explaining the assumptions, it is useful to discuss a simple 
example. Consider a two-photon experiment, where a quantum state should 
be analysed by performing state tomography. For that, Alice and Bob
have to measure all the nine possible combinations of the Pauli matrices 
$\sigma_i \otimes \sigma_j$ for $i,j \in \{x,y,z\}$. In practice, this
can be done as follows: Alice and Bob measure the three 
Pauli matrices $\sigma_x, \sigma_y, \sigma_z$ by measuring the polarization 
in different directions, getting the possible results $+$ and $-$. These results correspond to the projectors on the eigenvectors of the 
observable. By combining the results, they obtain one of four possible 
outcomes from the set $\{++,+-,-+,--\}$. The measurement is repeated $N$ 
times on copies of the state, where the outcome $++$ occurs $N_{++|ij}$ times 
etc. From that, one can obtain the relative frequencies $F_{++|ij}=N_{++|ij}/N$ 
and estimate the expectation values as $\mean{\sigma_i \otimes \sigma_j}_{\rm exp} = 
(N_{++|ij}-N_{+-|ij}-N_{-+|ij}+N_{--|ij})/N.$ In addition, the expectation values of
the marginals $\sigma_i \otimes \sigma_0$ (here and in the following, we set
$\sigma_0 =\eins$) can be determined from the same data. Given all the experimental 
results, Alice and Bob may then reconstruct the quantum state via the formula
\be
\hat \vr = \frac{1}{4}  \sum_{i,j \in \{0,x,y,z\}}\!\!\! \lambda_{ij} \sigma_i \otimes \sigma_j,
\mbox{  where  } \lambda_{ij} = \mean{\sigma_i \otimes \sigma_j}_{\rm exp}.
\label{eqlininv}
\ee
This simple quantum state reconstruction scheme is often called linear 
inversion. It assumes that the observed frequencies equal the probabilities,
we will discuss its advantages and disadvantages below. For the moment, we 
just use it as an example to illustrate the definitions and discussion 
concerning  systematic errors in experiments. 

Now we can formulate the assumptions that lead to the statistical model 
typically used in quantum experiments. We consider a scenario where one 
actively chooses between different measurements (e.g., the 
$\sigma_i \otimes \sigma_j$), each having a finite number of results. 
We use the label $s$ to denote the measurement setting and $r$ to denote 
the result. It is important to note that, if in an experiment using 
the setting $s$ one registers the result $r$, then this outcome $r|s$ 
is not just treated as a classical result. In addition, each outcome 
is tied to an operator $M_{r|s}$ (e.g., the projectors onto the 
eigenstates corresponding to the results $\{++,+-,-+,--\}$ of $\sigma_i \otimes \sigma_j$) that serves as the object to compute 
probabilities within quantum mechanics: If the underlying quantum state 
is characterized by the density operator $\varrho$, then the probability 
to observe $r|s$ is given by $P(r|s; \varrho)=\tr(\varrho M_{r|s})$. 
Therefore, this quantum mechanical description is one of the
essential ingredients to connect the observed samples with the parameters 
of the system that one likes to infer. Knowledge about this 
description can come from previous calibration measurements or from other 
expertise that one has acquired with the equipment. But one thing should 
be obvious: If one assumes a description $M_{r|s}$, which deviates from the 
true description in the experiment $\tilde M_{r|s}$, then things can go 
terribly wrong and these type of errors are the ones that we like to 
address in the following. 

Clearly, such deviations are presumably present in any model, but they are usually assumed to be small. However, considering the 
increased complexity of present experiments, one can ask 
the question, whether or not these deviations show up significantly 
in the data. Well known examples, like different detection efficiencies 
or dark-count rates in photo-detectors or non-perfect gate fidelities
for single-qubit rotations preceding the readout of a trapped ion,
could support this scepticism. However, these effects are hardly ever 
considered in the description of $M_{r|s}$.

Let us complete the list of assumptions. Most often each experiment of 
setting $s$ is repeated $N$ times, which are assumed to be independent 
and identically distributed trials. So one further assumes that one 
always prepares the same quantum state $\varrho$, measures the same 
observables $M_{r|s}$, and that both are completely independent~\footnote{This 
means that both, measurements and states are described by the corresponding 
$N$-fold tensor products. While such a property can be inferred for the states 
with the help of the de Finetti theorem~\cite{renner07a}, one should be aware 
that its exchangeability requirements do not apply to experiments where one 
actively measures first all the $s=1$ measurements, followed by all $s=2$ 
measurements and so on.}. 
Clearly, also in all these steps there can be errors, for instance, due to 
drifts in the measuring devices or dead-times in detectors coming from 
previous triggering events. However, if everything works as planned, then 
it is not necessary anymore to keep track of the individual measurement 
results, since every information that can be inferred about the state 
parameters is already included in the count rates $N_{r|s}$ of the 
individual measurement results $r|s$. Their probability is then given 
by a multinomial distribution ${\rm Mult}[N, P(r|s;\varrho)]$ for each
setting, which
is the distribution characterizing $N$ repetitions of independent trials.
Here, the single event probabilities $P(r|s;\varrho)=\tr(\varrho M_{r|s})$ 
are calculated according to quantum mechanics and these are the only 
parameters that depend on the quantum state. 

Finally, the whole collection of distributions 
for all measurement settings is the exact parametric model used 
for most quantum experiments. These distributions are given by
the set
\begin{equation}
\label{eq:P_QM}
\mathcal{P}_{\rm QM} = \Big\{ P(\{ N_{r|s} \}_{r,s}; \varrho) 
= \prod_s {\rm Mult} \left[ N, \tr(\varrho M_{r|s}) \right], 
\mbox{ for all }\varrho \mbox{ with }\varrho~\geq~0, \tr(\varrho) = 1 \Big\},
\end{equation} 
and the observed probabilities are assumed to be an element of this set. 
In the following, we discuss how the validity of this model can be tested.

\subsection{Testing the assumptions}
How can one test in this framework whether the assumed 
measurement description is correct for the experiment? As 
a first try, we could intersperse the experiment with test 
measurements, in which one prepares previously characterized
states. But such an option seems very cumbersome, independent
of problems like how to characterize the test states in the 
first place and to ensure that they are well prepared in 
between the true experiment. In contrast, we want to do it
more directly and this becomes possible, at least partially, 
by exploiting that quantum states only allow a restricted set
of event probabilities.

Let us first discuss the idea for the case where one has access to
the true event probabilities $P_0(r|s)$ which can be attained from 
the relative frequencies $F_{r|s} = N_{r|s}/N \to P_0(r|s)$ 
in the limit $N \to \infty$. We want to know whether these observed 
probabilities are at all compatible with the assumed measurement 
description. This boils down to the question whether there exists 
a quantum state $\varrho_0$ with $P_0(r|s) = \tr(\varrho_0 M_{r|s})$ 
for all $r,s$. Since quantum states must respect the positivity 
constraint $\varrho \geq 0$, not all possible probabilities are 
accessible: For instance, if one measures a qubit along the $x,y,z$ 
directions, its corresponding probabilities will be constrained 
by the requirement that the Bloch vector must lie within the Bloch 
ball. To make this more general, assume that we have a certain set 
of numbers $w_{r|s}$ such that the observable 
$\sum_{r|s} w_{r|s} M_{r|s} \geq 0$ has no negative eigenvalues and
is, therefore, positive semidefinite. If the probabilities  $P_0(r|s)$ 
can indeed by realized by a quantum state, one has
\begin{equation}
\label{eq:ConsistencyWitness}
w \cdot P_0 \equiv \sum_{r,s} w_{r|s} P_0(r|s) 
= \sum_{r,s} w_{r|s} \tr(\varrho_0 M_{r|s})
= \tr[\varrho_0 \big(\sum_{r,s} w_{r|s} M_{r|s} \big)] \geq 0, 
\end{equation}
where the inequality holds because both operators are positive
semidefinite. Thus, if everything is correct one must get a 
non-negative value for $w \cdot P_0 \geq 0$. Consequently, 
whenever one observes $w \cdot P_0 < 0$, one knows that something 
must be wrong and that  the description of the measurements $M_{r|s}$ 
has some flaws. This type of inequalities is similar in 
spirit to Bell inequalities for local hidden variable 
models or entanglement witnesses for 
separable states \cite{hororeview, tgreview}. Let us point out that 
the above inequalities are necessary and sufficient. So, 
indeed any $P_0(r|s)$ which cannot originate from a 
quantum state, can be detected by appropriately chosen 
coefficients $w_{r|s}$ by $w \cdot P_0 < 0$~\cite{moroder08a}. 
Finally, we add that besides the positivity, some other constraints 
for the measurement description are conceivable. For instance, in 
the example of state tomography from above, the marginal 
$\mean{\sigma_x \otimes \sigma_0}$ should not depend on whether 
it has been derived from the measurement $\sigma_x \otimes \sigma_x$ 
or $\sigma_x \otimes \sigma_y$. This can be formulated as a linear 
dependency of the form
$\sum_{r,s} w_{r|s} M_{r|s} = 0$ and the corresponding constraint
even becomes an equality $w \cdot P_0 = 0$. 
 
Note that, with this test we ask the question whether the 
data $P_0(r|s)$ {\it fit at all} to the assumed measurement 
model $M_{r|s}$. But it should be clear that this approach can never 
serve as a proof that everything is correct in the experiment. 
For example, one can consider again the Bloch ball, where the measurement
model assumes perfectly aligned measurements in the $x,y,z$ directions, 
but in the true experiment one measures in slightly tilted directions 
which distorts the resulting Bloch ball. All states from this tilted 
Bloch ball, which lie outside the standard sphere, will be detected by 
the above method as being incompatible with the assumed model. For all 
other states, however, we do not see the difference because they are still
consistent with the model. 

Finally, let us address the point that we only collect count
rates in the  experiment. Since the relative frequencies 
$F_{r|s}$ are only approximations to the true probabilities $P(r|s)$, 
it is clear that a similar inequality as Eq.~\eqref{eq:ConsistencyWitness} 
does not need to hold anymore for $w \cdot F$, even if 
everything is correct. One would expect, however, that larger 
negative values are much less likely. This is indeed the case and
is made more quantitative via  Hoeffding's inequality \cite{hoeffding}.

This inequality states the following: Consider $N$ independent, 
not necessarily identically distributed, bounded random variables 
$X_i\in [a_i, b_i].$ Then the sample mean $\bar X= \sum_i X_i /N$ 
satisfies
\be
{\rm Prob}[ \bar X - \mathbbm{E}(\bar X) \leq -t] \leq \exp\Big(\frac{-2t^2N^2}{\sum_i (b_i-a_i)^2}\Big)
\ee
for all $t > 0$, where $\mathbbm{E}(\bar X)$ denotes the expectation value of 
$\bar X$. 
In practice, the main statement of this inequality is that for 
$N$ independent repetitions of an experiment, the probability of deviations 
from the mean value by a difference $t$ scales like $\exp(-t^2 N).$
It is important to stress that this result uses no extra assumptions, 
like $N$ being large, at all.

For our case, we can use Hoeffding's inequality to bound the probability
of observing data that violate positivity constraints as in
Eq.~(\ref{eq:ConsistencyWitness}). More precisely, we can derive the following 
statement \cite{moroder13a}: For all distributions compatible with quantum mechanics,
the probability 
to observe frequencies $ \{F_{r|s}\}_{r,s}$ such that 
$w \cdot F < -\varepsilon$ is bounded by 
\begin{equation}
\label{eq:Hoeffding}
{\rm Prob}_{P} \big[ w \cdot F < - \varepsilon \big] 
\equiv \sum_{ \stackrel{\{ n_{r|s} \}_{r,s}:}{w \cdot F < -\varepsilon}}
P( \{ n_{r|s} \}_{r,s}) \leq \exp \Big(-\frac{2 \varepsilon^2 N}{ C_w^2} \Big), 
\mbox{ for all } P \in \mathcal{P}_{QM}
\end{equation} 
with $C_w^2 = \sum_s (w_{{\rm max}|s} - w_{{\rm min}|s})^2$ 
and $w_{{\rm max}|s}$, $w_{{\rm min}|s}$ being the extreme 
values of $\{ w_{r|s} \}_r$. Again, this can be interpreted as
showing that if everything is correct, then the probability of 
finding a violation of the positivity constraint is exponentially
suppressed.

We can use this statement as follows: Suppose that we 
should reach a conclusion whether the observed data are 
``compatible'' or ``incompatible'' with our assumed model. 
Of course, if we say ``incompatible'', we do not want to 
reach this conclusion too often, if indeed everything is 
perfect. For definiteness, we may assume that the probability 
of claiming incompatibility if everything is correct
should
be at maximum $\alpha = 1 \%$. We then use Eq.~\eqref{eq:Hoeffding} 
to deduce the threshold value that we need to beat, 
$\varepsilon_{\alpha} = \sqrt{ C_w^2 |\log(\alpha)| / 2 N}$. 
If we now carry out the experiment and register click rates 
with $w \cdot F < -\varepsilon_{\alpha}$, we know 
that there was at most a $\alpha = 1 \%$ chance that we 
would have registered such badly looking data, if everything 
is correct. Since this would be really bad luck we would 
rather say  ``incompatible'', and assume that some systematic error was present~\footnote{Since one typically 
likes to leave the choice of appropriate levels of $\alpha$ 
to the reader one can also report the p-value~\cite{mood} 
of the observed data: It is the smallest 
$\alpha$ with which we would have still said ``incompatible'' 
with the test.}.

In practice, this test can be used to detect systematic errors
in various scenarios: In ion trap experiments, a typical systematic
error comes from the cross talk between the ions, i.e. the fact that a
laser focused on one ion also influences the neighbouring ions. This phenomenon can  be detected with the presented method \cite{moroder13a}.
The second application are Bell experiments: In these experiments, the choice of the measurements on 
one party should ideally not influence the results of the other party and
a violation of this condition completely invalidates the result of a Bell 
test. Again, this non-signalling condition can be formulated as linear constraints on the probabilities and this can be tested with the presented method. In all these applications, the determination of the vector $w$
characterizing the positivity constraint or the linear constraint can be
done as follows: One splits the observed data into two parts. From the 
first part one determines the $w$ leading to the maximal violation of
the respective constraint for the first half. Then, one applies this $w$
as a test to the second part of the data. If the violations of the constraint
are only due to statistical fluctuations, the respective $w$ for the two
parts of the data are uncorrelated and the test will not find a significant 
violation of the constraint.

Let us point out that the mathematical framework just 
described is called a hypothesis test~\cite{mood}, in 
which one tests the null-hypothesis $N_0$: ``compatible'', 
against the alternative $A$: ``incompatible''. The special 
property of such a test is that there is an asymmetry about 
the two types of errors that can occur. As already explained, 
our concern is that, when saying ``incompatible'', then this statement 
is more or less correct.  The other error can occur when we
respond ``compatible'' to incompatible data.
Naturally, this error characterizing the
detection strength of our test, ideally, should also be made small.
However, it is not possible to reduce both errors equally simultaneously.
Nevertheless, since we cannot detect 
all possible systematic deviations from  the 
assumed model, anyway, one should not be too euphoric about the statement ``compatible'' in this sense. 

Note that, while the presented test has been build 
up by first deriving specific inequalities for event 
probabilities and then equipping it with the necessary 
statistical rigour to arrive at an hypothesis test, one
can also take the other direction, by using techniques which 
are known to be good for hypothesis tests and apply them
to the special statistical model of the quantum experiments.
We have done this for the so-called generalized likelihood-ratio 
test~\cite{mood} and details can be found in Ref.~\cite{moroder13a}. 
Finally, other tests for systematic errors can be found in 
Refs.~\cite{schwarz11a,langford12a,van_enk13a}.

\section{Performing state tomography}
In the previous section, we have seen that care has to be taken when making the
measurements on the quantum system. In this section, we show that the
interpretation of tomographic data, such as the reconstruction of the quantum
state, has to be done with care, too.  Otherwise, one introduces yet another
class of systematic errors.

\subsection{Problems with state estimates}
We are used to summarize experimental data by an estimate together 
with an error margin. In quantum state tomography this corresponds 
to an estimate for the density matrix together with an error region. 
So, the first question is how one can obtain an estimate $\hat\vr$ 
for the experimentally prepared density matrix $\vr_{\exp}$ from 
the observed frequencies $F_{r|s}.$
The simplest approach is to use linear inversion, that is, the method 
given in Eq.~(\ref{eqlininv}). This has, however, 
at first sight some disadvantages: Due to statistical fluctuations
the observed frequencies are not equal to the true probabilities and 
this leads to the consequence that the reconstructed ``density matrix''
will typically have some negative eigenvalues. This makes the further
analysis of the experiment, e.g. the evaluation of entanglement measures,
not straightforward. In order to circumvent this, one often makes a density 
matrix reconstruction by setting
\be
\hat\vr = \arg \max_{\sigma \geq 0} \FF(F_{r|s},\sigma).
\ee
Here, one optimizes a target function $\FF$ over all density matrices
$\sigma$ and the optimal $\sigma$ will obviously be a valid density matrix. Examples for
this type of state reconstruction are the maximum-likelihood reconstruction or the
least-squares reconstruction, both are frequently used for experiments in quantum 
optics.

An important property of such an estimator is the question whether it is biased
or unbiased. This  means the following:  The underlying state $\vr_{\rm exp}$
leads via the multinomial distribution to a probability distribution over the
frequencies $F_{r|s}$. The estimator $\hat \vr$ is a function from the observed data (the 
frequencies $F_{r|s}$) to the state space. In this way, the original state $\vr_{\rm exp}$
induces a probability distribution over the estimators $\hat \vr$, and one can ask whether
the expectation value of this equals the original state, $\mathbbm E[\hat\varrho] 
\stackrel{?}{=} \varrho_\text{exp}.$ If this is the case, the estimator is unbiased, 
otherwise it is biased. It must be stressed, however, that biased estimators are not
necessarily useless or bad, as it all depends on the purpose the estimator is used for. 

\begin{figure}[t]
\begin{center}
\includegraphics[width=9cm]{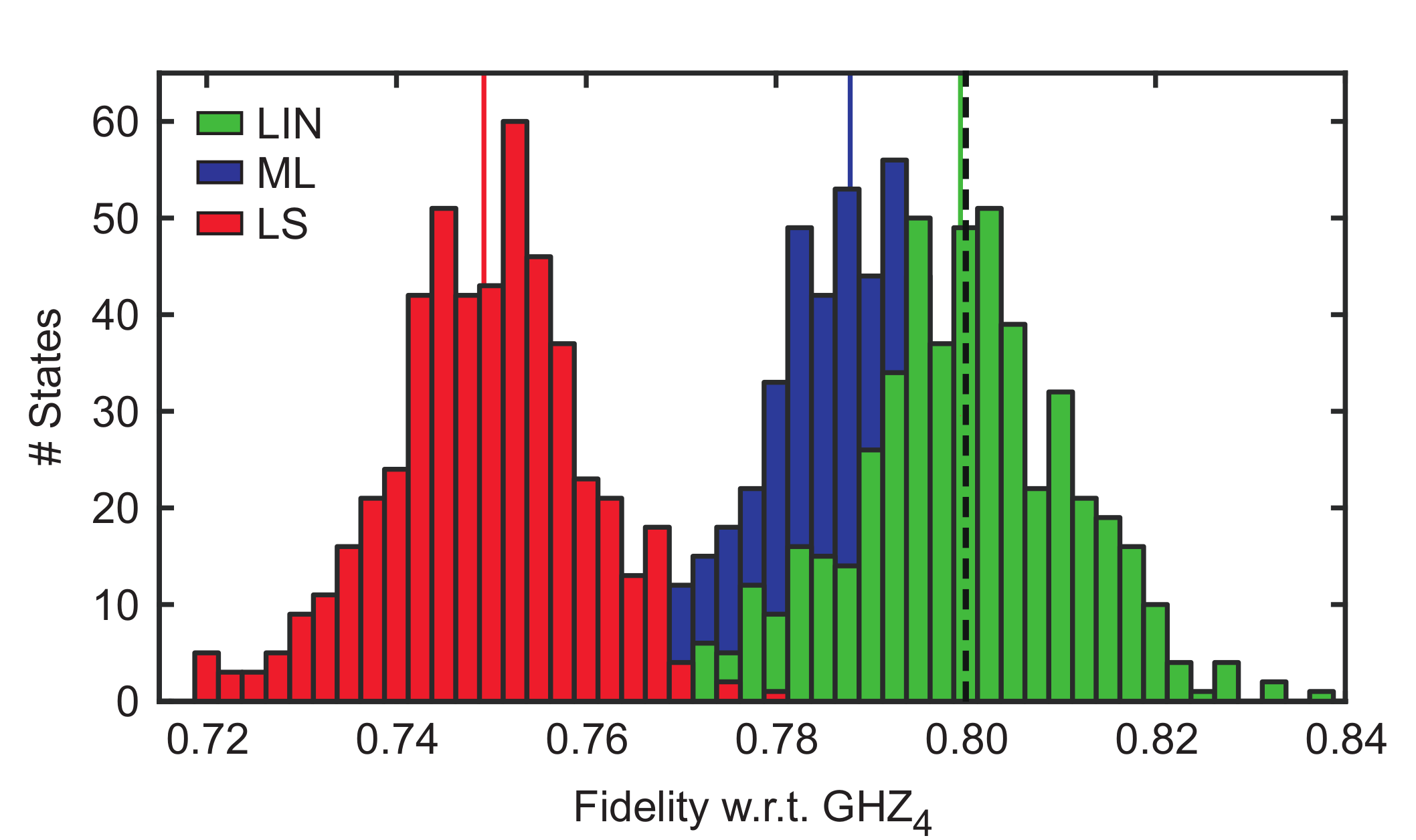}
\end{center}
\caption{Demonstration of the bias for different state estimators. A given state 
$\vr_{\rm exp}$ having 80\% fidelity with the GHZ state was used to sample the distribution
of the estimator $\hat \vr$ in state space, and from these samples the fidelities with
the GHZ state were computed. The maximum-likelihood (ML) and least-square (LS) estimators
clearly  underestimate the fidelity, while the linear inversion (LIN) is unbiased.
The figure is taken from Ref.~\cite{biasschwemmer}.
\label{schwemmerbild}}
\end{figure}

For quantum state reconstruction one can prove the following: Any
state reconstruction scheme that yields a density matrix from experimental 
data will be biased, i.e., on average, the reconstructed state $\hat\varrho$ 
will not be the state used in the experiment, $\mathbbm E[\hat\varrho] \ne
\varrho_\text{exp}$.  A proof of this statement was given in
Ref.~\cite{biasschwemmer}, but the following example demonstrates that 
the problem in finding an unbiased estimator comes from the fact that
the quantum mechanical state space is bounded by the positivity constraint.
Consider a coin toss where we are interested in the modulus of the difference 
between the probability of obtaining heads or tails, $\Delta=
|p_h-p_t|= 1-2 \min\{ p_h, 1-p_h\}$. This quantity cannot be negative, so also 
an estimator should not be negative. Let us assume that an estimator $\hat \Delta$ 
is unbiased, $\mathbbm E[\hat\Delta] = \Delta_\text{exp}$. Then, for any experimental 
data that could come from a fair coin ($\Delta_\text{exp}=0$) we cannot have 
$\hat\Delta > 0$ since this would imply $\mathbbm E[\hat\Delta] = 
\sum_k (1/2)^{n} (\begin{smallmatrix}n\\k\end{smallmatrix})\hat\Delta(k) > 0$,
where $k$ denotes the number of occurrences of heads. On the other hand, any 
possible number of heads and tails is compatible with a fair coin. 
So, the estimate $\hat \Delta$ for \emph{any} data must be 0. Then in particular 
$\mathbbm E[\hat\Delta] = 0$, which means that $\hat \Delta$  is a biased estimator
whenever the coin is {\it not} 
fair.

Apart from this theoretical argument, the question arises whether this effect
plays a significant role in practical quantum state reconstruction.
Unfortunately, this is the case and this effect can causes substantial fidelity
underestimation or spurious entanglement detection in realistic scenarios
\cite{biasschwemmer}. This problem applies to the established schemes for
reconstructing a density matrix, in particular to the maximum-likelihood method \cite{hradil} and the constrained least-squares method \cite{leastsquares}.
So, how
large is the bias? For example, in a tomography of a four-qubit GHZ state with
fidelity $0.8$, when reconstructing from a total number of {8100} samples, the
state from a maximum-likelihood estimate has a fidelity of $0.788\pm0.010$
\cite{biasschwemmer}, i.e., the fidelity is systematically underestimated 
(see also also Fig.~\ref{schwemmerbild}). Such an underestimation may be considered 
to be unfortunate, but acceptable. However, it was also demonstrated that 
maximum-likelihood and least-square methods tend to {\it overestimate} the 
entanglement. In fact, for a clearly separable state the reconstructed states 
can be always entangled, thus leading to spurious entanglement detection 
\cite{biasschwemmer}. This is not acceptable for many experiments.

A way to avoid the bias is to accept that the reconstructed density matrix 
is not always a valid quantum state and can have negative eigenvalues. The
simplest unbiased method is linear inversion explained above. More generally,
if $M_{r|s}$ are the operators corresponding to the measurement outcomes in 
a complete tomographic measurement scheme, then one can find operators $X_{r|s}$ 
such that for all states $\varrho= \sum_{r|s} X_{r|s} P(r|s;\varrho)$ 
holds, generalizing Eq.~\eqref{eqlininv} \footnote{The new operators 
$X_{r|s}$ may be necessary, since the $M_{r|s}$ can be overcomplete or
not orthogonal.}. The estimate given by linear reconstruction 
is then $\hat\varrho= \sum_{r|s} X_{r|s} F_{r|s}$, where $F_{r|s}$ are the relative 
frequencies of the result $r$ for setting $s$. This estimate is unbiased but it comes 
with the price that in all realistic scenarios $\hat\varrho$ has some negative 
eigenvalues and hence it is not a valid density matrix. Depending on the intended 
use of the reconstructed density matrix this may be problematic, but it was shown in 
Ref.~\cite{biasschwemmer} that entanglement measures or the Fisher information can still
be estimated. In addition, we stress that the 
eigenvectors corresponding to these negative eigenvalues are randomly distributed 
in the following sense: If we choose a rank one projection $\ketbra{\alpha}$ 
independently of the data, then the probability that $\tr(\hat\varrho\ketbra{\alpha})<-\epsilon$ 
is exponentially suppressed, as can be seen from the inequality in~\eqref{eq:ConsistencyWitness}.

\subsection{Problems with error regions}
Any report of an experiment has to equip the reported results with error 
bars. In the case of a density matrix, this will be a high-dimensional 
error region. When specifying an error region, one first has to decide 
between the Bayesian framework and the frequentistic framework. A Bayesian 
analysis gives a credible region, which has the property that with high 
probability the actual state is in this region. A frequentist's analysis 
gives a confidence region, which is a map from the data to a region in 
state space such that with high probability the region contains the 
actual state. There is a long debate in mathematical statistics which 
method is appropriate, but most of the subsequent discussion is 
independent of this dispute.

Before discussing the advantages and disadvantages of an error region, it is
important to remember, that the variance does in general not give an
appropriate error region. This occurs in particular if the underlying
distributions are far from being Gaussian. But for state tomography, the data
is sampled from a multinomial distribution, typically with a very low number of
events. Indeed, in many experiments the number of clicks per measurement
outcome is about ten, but sometimes even below one. Also the method of
bootstrapping may  yield an inappropriate error region. In bootstrapping, one uses an
estimate $\hat\varrho$ for the state (parametric bootstrapping) or the
empirical distributions of the outcomes of the measurements $F_{r|s}$
(non-parametric bootstrapping) in order to estimate the variance of the
estimate. This estimate is usually obtained by Monte Carlo sampling from the
corresponding distributions. There is no particular reason that this should be
a good error region, and it was also demonstrated that the most commonly used
schemes yield invalid error regions.

Methods to obtain valid error regions both in the Bayesian \cite{credible} and 
in the frequentistic framework \cite{freqregion} have been suggested, however, they 
turn out to be notoriously difficult to compute. But even when it is possible 
to achieve a proper error region, one has to keep in mind that the size of 
the error region scales with the dimension of the underlying Hilbert space, i.e.,
exponentially with the number of qubits. This makes it very difficult to perform
state tomography of a large system with a reasonable sized error region. Fortunately, 
in many situations the error region for the state is not of uttermost importance. 
Often one is only interested in certain scalar quantities like a measure of
entanglement or the fidelity with the (pure) target state. In this cases it is 
possible to infer an appropriate confidence region directly from the data, 
without taking the detour over an error region for the density operator. 
This is particularly simple, if the quantity of interest is linear in the 
density matrix, e.g., the fidelity with a pure state $F=\langle\psi|\varrho|\psi\rangle$. 
One can again use Hoeffding's tail inequality in order to obtain a lower 
bound $\hat F_l$ on the fidelity. The promise is then that  $P(\hat
F_l>\langle\psi|\varrho_\mathrm{exp}|\psi\rangle) < 1\%$ for any state
$\varrho_\mathrm{exp}$.  A general method to provide such confidence regions
for convex functions, like the bipartite negativity or the quantum Fisher
information, has been introduced in Ref.~\cite{biasschwemmer}.

\section{Analysing density matrices}
In the last section of this article, we assume that a valid multiparticle 
density matrix $\vr$ is given and the task is to analyse its properties. 
 Naturally, many questions can be asked about
a density matrix, but we concentrate on two of them.
First, we consider the question whether the 
state is genuinely multiparticle entangled or not. We explain a powerful 
approach for characterizing multiparticle entanglement with the help of 
so-called PPT mixtures and semidefinite programming. Second, we 
consider the problem of characterizing the complexity of a given quantum 
state and explain an approach using exponential families. For example, 
in this approach a state that is a thermal state of a Hamiltonian with 
two-body interactions only, is considered to be of low complexity and the 
distance to these thermal states can be considered as a measure of complexity. 
The underlying techniques also allow to characterize pure states which are 
not ground states of a two-body Hamiltonian.

\subsection{Characterizing entanglement with PPT mixtures}

\subsubsection{Notions of entanglement ---}
Before explaining the characterization of multiparticle entanglement, 
we have to explain some basic facts about entanglement on a two-particle 
system. The definition of entanglement is based on the notion of local 
operations and classical communication (LOCC). If a quantum state can 
be prepared by LOCC, it is called separable, otherwise it is entangled. For pure
states, this just means that product states of the form $\ket{\phi} = \ket{\alpha}
\otimes \ket{\beta}$ are separable and all other states (e.g. the singlet state
$\ket{\psi^-}=(\ket{01}-\ket{10})/\sqrt{2}$) are entangled. 
If mixed states are considered, a density matrix $\vr$ is separable, if it can be 
written as a convex combination of product states,
\be
\vr=\sum_k p_k \vr_A^k \otimes \vr^k_B,
\label{entdef}
\ee
where the $p_k$ form a probability distribution, so they are non-negative 
and sum up to one. Physically, the convex combination means that Alice 
and Bob can prepare the global state by fixing the joint probabilities
with classical communication and then preparing the states $\vr^A_k$ 
and $\vr^B_k$ separately. The question whether or not a given quantum 
state is entangled is, however, in general difficult to answer. This 
is the so-called separability problem 
\cite{hororeview, tgreview}.

Many separability criteria have been proposed, but none of them 
delivers a complete solution of the problem. The most famous 
separability test is the criterion of the positivity of 
the partial transpose (PPT criterion) \cite{pptcriterion}.  For that, one considers 
the  partial transposition of a density matrix 
$\vr=\sum_{ij,kl} \vr_{ij,kl}\ket{i}\bra{j} \otimes \ket{k}\bra{l}$, 
given by
\be
\vr^{T_A}=\sum_{ij,kl} \vr_{ij,kl}\ket{j}\bra{i} \otimes \ket{k}\bra{l}.
\ee
In an analogous manner, one can also define the partial transposition
$\vr^{T_B}$ with respect to the second system. The PPT criterion states 
that for any separable state $\vr$ the partial transpose $\vr^{T_A}$,
(and consequently also $\vr^{T_B}=(\vr^{T_A})^T$) has no negative eigenvalues and is therefore positive semidefinite. So, 
if one finds a negative eigenvalue of $\vr^{T_A}$, then the state $\vr$ must 
necessarily be entangled. The PPT criterion solves the separability problem 
for low dimensional systems (that is, two qubits or one qubit and one qutrit) 
\cite{hororeview}, but in all other cases the set of separable states is a strict subset 
of the PPT states. The entangled states which are PPT are of great theoretical 
interest: It has been shown that their entanglement can never be distilled to 
pure state entanglement, even if many copies of the state are available. This 
weak form of entanglement is then also called {\it bound entanglement} and bound 
entangled states are central for many challenging questions in quantum information 
theory.

The characterization of entanglement becomes significantly more complicated, 
if more than two particles are involved. Let us consider three particles 
(A, B, C). First, a state can be fully separable, meaning that it does not contain
any entanglement and is of the form 
$\ket{\phi^{\rm fs}}=\ket{\alpha}\otimes\ket{\beta}\otimes\ket{\gamma}$. If a state is entangled, 
one can further ask whether only two parties are entangled or all three parties. 
For instance, in the state $\ket{\phi^{\rm bs}}=\ket{\psi^-}_{AB}\otimes \ket{\gamma}_C$ 
the parties A and B are entangled, but C is not entangled with A or B, therefore the 
state is  called biseparable. Alternatively, if all parties are entangled with each 
other, the state is called genuine multipartite entangled \cite{tgreview}. For the simplest 
case of three two-level systems (qubits) it has been shown that even the genuine 
multipartite entangled states can be divided into two subclasses, represented
by the GHZ state $\ket{GHZ}= (\ket{000}+\ket{111})/\sqrt{2}$ and the W state
$\ket{W}=(\ket{001}+\ket{010}+\ket{100})/\sqrt{3}$. These subclasses are 
distinguished by the fact that a single copy of a state in one class cannot be 
converted via LOCC into a state in the other class, even if this transformation 
is not required to be performed with probability one \cite{tgreview}. 

The classification of entanglement for pure states can be extended to 
mixed states by considering convex combinations as in Eq.~(\ref{entdef}). 
First, a mixed state is fully separable, if it can be written as a convex 
combination of  fully separable states
\be
\vr^{\rm fs}=\sum_k p_k \vr^k_A \otimes \vr^k_B \otimes \vr^k_C,
\ee
and a state is biseparable, if it can be written as a mixture of 
biseparable states, which might be separable with respect to different 
partitions,
\be
\vr^{\rm bisep} = p_1 \vr^{\rm sep}_{A|BC} + p_2 \vr^{\rm sep}_{B|AC} + p_3 
\vr^{\rm sep}_{C|AB}.
\label{mpentdef}
\ee
The different notions of entanglement in the multipartite case 
and the different bipartitions that have to be taken into account
imply that the question whether a given mixed multipartite state 
is entangled or not is extraordinarily complicated. 

\begin{figure}[t]
\begin{center}
\includegraphics[width=7cm]{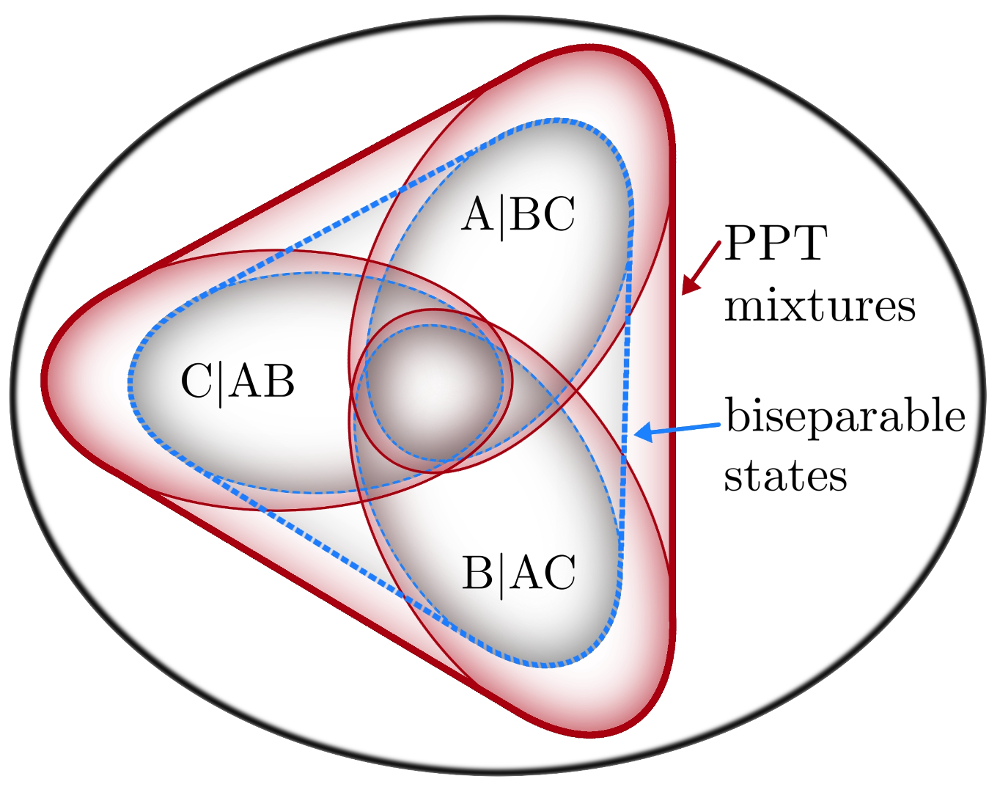}
\end{center}
\caption{Schematic view of the states which are PPT mixtures and the 
biseparable states for three particles. There are three possible 
bipartitions, and the corresponding sets of states which are separable 
or PPT for the bipartition. The figure is taken from Ref.~\cite{PPTMixer}.
\label{pptmixfig}}
\end{figure}

\subsubsection{The approach of PPT mixtures ---}
A systematic approach for characterizing genuine multiparticle 
entanglement makes use of so-called PPT mixtures \cite{PPTMixer}. Instead 
of asking whether a state is a mixture of separable states with 
respect to different partitions as in Eq.~(\ref{mpentdef}), one 
asks whether it is a mixture of states which are PPT for the 
bipartitions
\be
\vr^{\rm pptmix} = p_1 \vr^{\rm ppt}_{A|BC} + p_2 \vr^{\rm ppt}_{B|AC} + p_3 
\vr^{\rm ppt}_{C|AB}.
\label{pptmixdef}
\ee
Since the separable states are a subset of the PPT states, any biseparable 
state is also a PPT mixture. This means that if a state is no PPT mixture, 
then it must be genuine multipartite entangled (see also Fig.~\ref{pptmixfig}). 

At first, it is not clear what can be gained by this redefinition of the problem. 
First, the condition for PPT mixtures is a relaxation of the definition of 
biseparability and it might be that the conditions are relaxed too much, implying
that not many states can be detected by this method. Second, it is not clear how
the criterion for PPT mixtures can be evaluated in practice and whether this is
easier than evaluating the conditions for separability directly. In the following, 
however, we will see that the question whether a state is a PPT mixture or not can
directly be checked with a technique called semidefinite programming. Furthermore, 
the approximation
to the biseparable states is rather tight, and for many families of states the 
property of being a PPT mixture coincides with the property of being biseparable.

\subsubsection{Evaluation of the criterion ---}
Let us  discuss the evaluation of the condition for PPT mixtures. For that, we 
need to introduce the notion of entanglement witnesses. In the two-particle case, 
an entanglement witness $\WW$ is an observable with the property that the expectation 
value is positive for all separable states, $\tr(\vr^{\rm sep} \WW) \geq 0$. This implies that a measured negative 
expectation value signals the presence of entanglement. In this way, the concept
of an entanglement witness bears some similarity to a Bell inequality, where 
correlations are bounded for classical states admitting a local hidden variable 
model, while entangled states may violate the bound. 

How can entanglement witnesses be constructed? For the two-particle
case a simple method goes as follows: Consider an observable of the form
\be
\WW = P + Q^{T_A},
\label{twoparticlewitness}
\ee
where $P\geq 0$ and $Q \geq0$ are positive semidefinite operators.  
Using the fact that $Tr(X Y^{T_A}) = Tr(X^{T_A} Y)$ for arbitrary 
operators $X,Y$, we find that for a separable state 
$Tr(\WW \vr^{\rm sep}) = Tr(P \vr^{\rm sep}) + Tr (Q (\vr^{\rm sep})^{T_A}) 
\geq 0,$ since $\vr^{\rm sep}$ 
has to be PPT. Therefore, the observable $\WW$ is an entanglement witness, 
which may be used to detect the entanglement in states that violate the PPT
criterion. 

This construction can  be used to decide whether a given three-particle
state is a PPT mixture or not. For that, consider the optimization problem
\bea
\underset{\WW, P_i, Q_i}{\mbox{minimize}} &\quad& Tr (\vr \WW)
\nonumber
\\
\mbox{subject to:} & \quad&  \WW = P_1 + Q_1^{T_A} = P_2 + Q_2^{T_B}= P_3 + Q_3^{T_C}
\mbox{ and }
\nonumber
\\ && P_i \geq 0 \mbox{ for } i =1,2,3 \mbox{ and }
\nonumber
\\ && \eins \geq  Q_i \geq 0 \mbox{ for } i =1,2,3.
\label{pptmixsdp}
\eea
The constraints guarantee that the observable $\WW$ is of the form as in 
Eq.~(\ref{twoparticlewitness}) for any of the three bipartitions. This means, 
that if a state is a PPT mixture as in Eq.~(\ref{pptmixdef}), the expectation value
$ Tr(\vr \WW)$ has to be non-negative. On the other hand, one can show that if a state
is not a PPT mixture, then the minimization problem will always result in a strictly
negative value \cite{PPTMixer}. In this way, the question whether a state is a PPT mixture 
or not, can be transformed into a optimization problem under certain constraints.

The point is that the optimization problem belongs to the class of 
semidefinite
programs (SDP). An SDP is an optimization problem of the type
\bea
\underset{x_i}{\mbox{minimize}} &\quad& \sum_i c_i x_i
\nonumber
\\
\mbox{subject to:} & \quad& F_0 + \sum_i x_i F_i \geq 0,
\eea
where the $c_i$ are real coefficients defining the target function, the $F_i$
are hermitean matrices defining the constraints and the $x_i$ are real coefficients
which are varied. This type of optimization problem has two important features \cite{SDP}.
First, using the so-called dual problem one can derive a lower bound on the 
solution of the minimization, which equals the exact value under weak conditions. 
This means that the optimality of a solution found numerically can be demonstrated.
In this way, one can prove rigorously by computer whether a given state is 
a PPT mixture or not. Second, for implementing an SDP in practice there are 
ready-to-use computer algebra packages available and therefore the practical
solution of the SDP is straightforward.

\subsubsection{Results ---}
Concerning the characterization of PPT mixtures, the following results have been
obtained:
\begin{itemize}

\item First, the practical evaluation of the SDP in Eq.~(\ref{pptmixsdp}) can be 
carried out easily with standard numerical routines. A free ready-to-use package 
called {\tt PPTMixer} is available online \cite{pptmixerprogram}, and it solves 
the problem  for up to six qubits on standard computers. For a larger number of 
particles, the numerical evaluation becomes difficult, but analytical approaches
are also feasible \cite{PPTMixer, martinana}.

\item For many families of states, the approach of the PPT mixtures delivers 
the strongest criterion of entanglement known so far. For many cases it even
solves the problem of characterizing multiparticle entanglement. For instance, 
three-qubit permutation-invariant states are biseparable, if and only if they
are PPT mixtures \cite{pipptmixer}. The same holds for states with certain 
symmetries, like
GHZ diagonal states or four-qubit states diagonal in the graph-state basis 
\cite{martinana, PhysRevA.87.012322}.

\item Nevertheless, the approach of PPT mixtures can not detect all multiparticle
entangled states. 
There are examples of genuinely entangled three-qubit and three-qutrit states, 
which are PPT mixtures \cite{gezaroof, huberrb}. For an increasing dimension 
and number of particles one can even show that the probability that a given 
multiparticle entangled state can be detected by the PPT mixture approach 
decreases \cite{lancien}. This finding, however, is in line with the 
observation that also in bipartite high-dimensional systems no single entanglement 
criterion detects a large fraction of states \cite{beigi}.

\item The value $\NN(\vr) = - Tr(\vr \WW)$, that is, the amount of violation of the 
witness condition is a  computable entanglement monotone for genuine multiparticle 
entanglement \cite{martinana}. It can be called the genuine multiparticle negativity, as it generalizes the entanglement measure of bipartite negativity.

\item An interesting feature of the PPT mixer approach is that it can also be 
evaluated, if only partial information on the state $\vr$ is available. Namely, 
if only the expectation values $\mean{A_i}$ of some observables $A_i$ are known, 
one can add in the SDP in Eq.~(\ref{pptmixsdp}) that the witness $\WW$ should be
a linear combination of the measured observables $\WW= \sum_i \lambda_i A_i$.
It can be shown that this is then still a complete solution of the problem, meaning
that the SDP returns a negative value, if and only if all states that are compatible
with the data $\mean{A_i}$ are not PPT mixtures. 
\end{itemize}

\subsection{Characterizing the complexity of quantum states}
Besides the question whether a given multiparticle quantum state is entangled 
or not, one may also be interested in other questions about a reconstructed quantum 
state $\vr$. For instance, one may ask: Is the given state is a ground state 
or thermal state of a simple Hamiltonian? In the following, we will explain how 
this question can be used to characterize the complexity of a many-body quantum state.

\subsubsection{Exponential families ---}
First, one can consider the set of all possible two-body Hamiltonians. For multi-qubit
systems they are of the form
\be
H_2 = \sum_{i,\alpha} \lambda^{(i)}_\alpha \sigma^{(i)}_\alpha +
\sum_{i,j,\alpha,\beta} \mu^{(ij)}_{\alpha\beta} \sigma^{(i)}_{\alpha} 
\sigma^{(j)}_\beta + \nu \eins,
\ee
where $\sigma_\alpha^{(i)}$ is the Pauli matrix $\sigma_\alpha$ acting on
the $i$-th qubit. This Hamiltonian $H_2$ contains, apart from the identity,
single-particle terms and two-particle interactions. However, no geometrical 
arrangement of the particles is assumed and the 
two-particle interactions are between arbitrary particles and 
not restricted to  nearest-neighbour interactions. We also denote the set of all two-particle Hamiltonians by $\HH_2$, and in a similar manner one can define the sets of $k$-particle Hamiltonians $\HH_k$.

Given the set of $k$-particle Hamiltonians, we can define the so-called exponential
family of all thermal states 
\be
\QQ_k = \{ \exp\{H_k\} \mbox{ with } H_k \in \HH_k\},
\ee
where the normalization of the state has been included into the Hamiltonian
via the term $\nu \eins$.

If a given quantum state is in the family $\QQ_k$ for small $k$, then one can consider it to be less complex, since only a simple Hamiltonian with few 
parameters are required to describe the interaction structure. One the other 
hand, if a state is not in the exponential family, one can consider the distance 
\be
D_k(\vr||\QQ_k):=\inf_{\eta \in \QQ_k} D(\vr||\eta)
\ee
with $D(\vr||\eta) = \tr[\vr \log(\vr)]-\tr[\vr \log(\eta)]$ being the 
relative entropy, as a measure of the complexity of the quantum state.
The optimal $\eta$ is also called the information projection $\tilde\vr_k$ 
and one can show that this $\tilde\vr_k$  is the maximum likelihood approximation 
of $\vr$ within the  family $\QQ_k$ \cite{KOJA09}. Below, we will explain several 
further equivalent characterizations which can help to solve the
underlying minimization problem. 

This type of complexity measure has been first discussed for the case of 
classical probability distributions in the context of information geometry 
\cite{amari01}. The measure $D_1$  is also known as the multi-information in complexity
theory \cite{AyKnauf07}. For classical complex systems, these quantities have been used to study
the onset of synchronization and chaos in coupled maps or cellular automata \cite{KOJA09}. For the quantum case, this measure and its properties have been discussed 
in several recent works \cite{Zhou08, Zhou09, GaGu11, niekampjpa}.

At this point, it is important to note that in the quantum case as well as 
in the classical case the quantity $D_k$ does not necessarily decrease under 
local operations \cite{Zhou09, GaGu11}. Simple examples for this fact follow from observation
that taking a thermal state of a two-body Hamiltonian and tracing out one particle 
typically leads to a state that is not a thermal state of a two-body Hamiltonian
anymore. Therefore, the quantity $D_k$ should not be considered as a measure of
correlations in the quantum state, it is more appropriate to consider it as a measure of the complexity of the state.

\begin{figure}[t]
\begin{center}
\includegraphics[width=7cm]{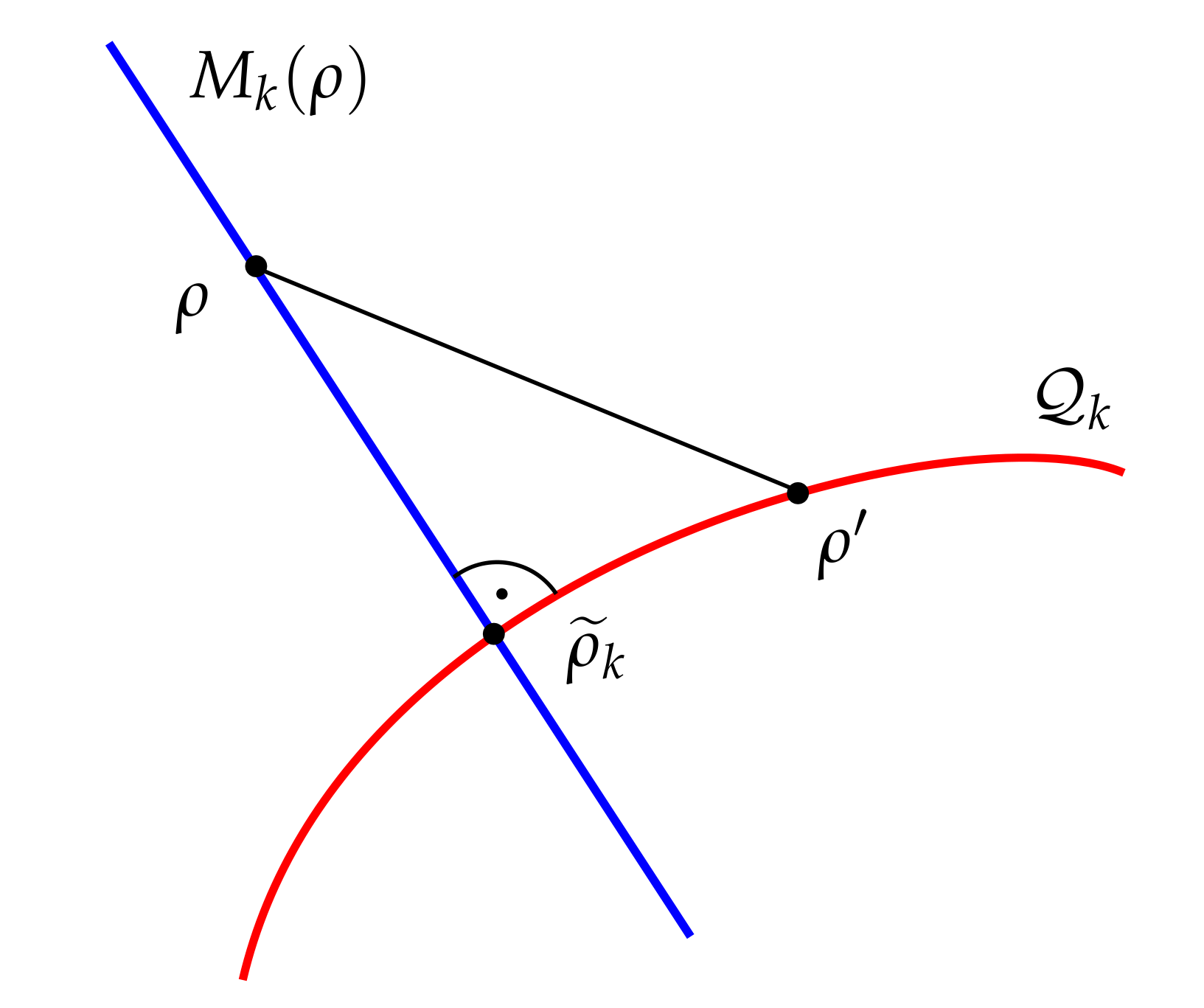}
\end{center}
\caption{The information projection $\tilde\vr_k$ of the state $\vr$
is the closest state to $\vr$ within the exponential family $\QQ_k$. 
$M_k$ denotes the set of all quantum states that have the same $k$-body 
marginals as $\vr$, and can also be used to characterize $\tilde\vr_k$.
For arbitrary states $\vr'$ within $\QQ_k$ the relation 
$D(\vr || \vr') = D(\vr || \tilde\vr_k) + D(\tilde\vr_k ||\vr')$ holds, 
which resembles the Pythagorean Theorem. The figure is taken from Ref.~\cite{niekampjpa}
\label{expfamfig}}
\end{figure}

\subsubsection{Characterizing the approximation ---}
For the characterization of the information projection $\tilde\vr_k,$
the following result is quite helpful \cite{Zhou09}. First, let $\vr$ be an 
arbitrary quantum state, and $\tilde\vr_k$ be the information projection
onto the exponential family $\QQ_k$. Furthermore, let $M_k$ be the set of
all quantum states that have the same $k$-body marginals as $\vr$. $M_k$ 
is, contrary to $\QQ_k$, a linear subspace of the space of all density matrices
(see Fig. \ref{expfamfig}). 
Then, the following statements are equivalent:

\begin{enumerate}
\item[(a)] The state $\tilde{\vr}_k$ is the closest state to 
$\vr$ in $\QQ_k$ with respect to the relative entropy.

\item[(b)] The state $\tilde{\vr}_k$ has the maximal entropy 
among all states in $M_k.$

\item[(c)] The state $\tilde{\vr}_k$ is the intersection 
$\QQ_k \cap M_k.$

\end{enumerate}

This  equivalence  can be used for many purposes. For example, it 
is useful for developing an algorithm for computing the information 
projection \cite{Zhou09b, niekampjpa}. Instead of minimizing the relative entropy as a 
highly nonlinear function over $\QQ_k$, one can  do the 
following: One optimizes over all states in $\QQ_k$ with the aim to make 
the $k$-body marginals the same as for the state $\vr$. The resulting 
algorithm converges well and allows the computation of the complexity measure
$D_k$ for up to six qubits \cite{niekampjpa}.

Second, from the equivalences it follows that the multi-information $D_1$  
can directly be calculated, since the closest state to $\vr$ in the family 
$\QQ_1$ is the product state $\tilde \vr_1 =\vr_1 \otimes \vr_2 \otimes ... \otimes \vr_N$ built out of the reduced single-particle density matrices of $\vr$.
Clearly, $\tilde \vr_1$ has the same marginals as $\vr$ and maximizes the 
entropy.

\subsubsection{A five-qubit example ---} As a final example, let us discuss how the notion of exponential families can help to characterize ground states of
two-body Hamiltonians. For that, consider the five-qubit ring-cluster state 
$\ket{R_5}.$ This state is defined to be the unique eigenstate fulfilling
\be
\ket{R_5} = g_i \ket{R_5},
\ee
where 
$g_1 = 
\sigma_x\sigma_z \eins \eins\sigma_z,
g_2 = 
\sigma_z\sigma_x\sigma_z \eins \eins,
g_3 = 
\eins\sigma_z\sigma_x\sigma_z \eins,
g_4 = 
\eins \eins\sigma_z\sigma_x\sigma_z,
$ and $
g_5 = 
\sigma_z \eins \eins\sigma_z\sigma_x.
$
Here, the tensor product symbols have been omitted.
After appropriate local transformations, the 
ring-cluster state can also be written as 
\be
\ket{R_5} = \frac{1}{\sqrt{8}}
\big[
\ket{00000}+ 
\ket{00110}- 
\ket{01011}+ 
\ket{01101}+ 
\ket{10001}- 
\ket{10111}+ 
\ket{11010}+ 
\ket{11100} 
\big].
\ee
The ring-cluster state is an example of a so-called graph state, and plays
an important role in quantum error correction as a codeword of the five-qubit Shor code. 
It was known before that the 
state $\ket{R_5}$ cannot be the unique ground state of a two-body Hamiltonian
\cite{luttmer}. This, however, leaves the question open whether it can be approximated by 
ground states of two-body Hamiltonians. For instance, for three qubits it was
shown that not all pure states are ground states of two-body Hamiltonians, but all
pure states can be approximated arbitrarily well by such ground states \cite{wootterslinpo}.

The characterization of the exponential families from the previous section 
can indeed help to prove that the state has $\ket{R_5}$ has finite distance to 
all thermal states of two-body Hamiltonians. For that, first note that 
the two-body marginals of the state $\ketbra{R_5}$ are all maximally mixed 
two-qubit states. Then, one can directly find states which have the same
two-body marginals, but their entropy is larger than the entropy the state 
$\ketbra{R_5}$. This last property is, of course, not surprising, since $\ketbra{R_5}$ has as a pure state the minimal possible entropy. According 
to the previous
section, this already implies that $\ketbra{R_5}$ cannot be the thermal 
or ground state of any two-body Hamiltonian. 

Furthermore, if an arbitrary state $\vr$ has a high fidelity with $\ket{R_5}$
then the two-body marginals will be close to the maximally mixed states, and
in addition the entropy of $\vr$ will be small. This implies that one can find 
again states with the same marginals and higher entropy. Using these ideas and some
detailed calculations one can prove that if a state fulfils 
\be
F= \bra{R_5} \vr \ket{R_5}
\geq \frac{31}{32} \approx 0.96875. 
\ee
then it cannot be a thermal state of a two-body Hamiltonian \cite{niekampdiss}. This shows that
the state $\ket{R_5}$ cannot be approximated by thermal states
of two-body Hamiltonians. In principle, this bound can also be used to prove
experimentally that a given state is not a thermal state of a two-body Hamiltonian.

\section{Conclusion}
In conclusion we have explained several problems occurring 
in the analysis of multiparticle quantum states, ranging 
from systematic errors of the measurement devices to the 
characterization of ground states of two-body Hamiltonians. 
We believe that several of the explained topics are important
to be addressed in the future. First, since the current experiments
in quantum optics are getting more and more complex, advanced statistical
methods need to be applied in order to reach solid conclusions. Second, 
the analysis of ground states and thermal states of simple Hamiltonians
is relevant for quantum simulation and quantum control, so direct characterizations would be very helpful. 

\section{Acknowledgements}
We thank Rainer Blatt, Tobias Galla, Bastian Jungnitsch, Martin Hofmann, 
Lukas Knips, Thomas Monz, S\"onke Niekamp,  Daniel Richart,  Philipp Schindler,  Christian Schwemmer, and Harald Weinfurter for 
discussions and collaborations on the presented topics. Furthermore, 
we thank Mariami Gachechiladze, Felix Huber, and Nikolai Miklin for 
comments on the manuscript.  This work has been supported by the EU
(Marie Curie CIG 293993/ENFOQI, ERC Starting Grant GEDENTQOPT, ERC Consolidator Grant 683107/TempoQ), 
the FQXi Fund (Silicon Valley Community Foundation), and the DFG (Forschungsstipendium KL 2726/2-1).

\end{document}